\documentclass[aps,pre,preprint,floatfix]{revtex4}

\usepackage{latexsym}
\usepackage[usenames]{color}
\usepackage{amsthm}
\usepackage{hyperref}

\usepackage{amsmath}    % need for subequations
\usepackage{amsfonts}  %note how statements can be commented out
\usepackage{amssymb}
\usepackage{graphicx}
\usepackage{slashed}
\usepackage{fouridx}
\usepackage{tikz}
\usetikzlibrary{calc}

\theoremstyle{plain} 
\theoremstyle{plain} 
\theoremstyle{plain}  
\theoremstyle{plain} 
\theoremstyle{plain}  
\theoremstyle{remark} 
\theoremstyle{plain} 
\theoremstyle{remark}

\newcommand\CROSS[1]{%
  \hbox{%
    \vbox{
      \hrule
      \kern2.5pt
      \hbox{$#1$\,\strut}
    }%
  \vrule
  }\mskip\thickmuskip
}

\tikzset{
solid node/.style={circle,draw,inner sep=1.5,fill=black},
hollow node/.style={circle,draw,inner sep=1.5}
}

\newlength{\arrowsize}  
\pgfarrowsdeclare{biggertip}{biggertip}{  
  \setlength{\arrowsize}{0.5pt}  
  \addtolength{\arrowsize}{.5\pgflinewidth}  
  \pgfarrowsrightextend{0}  
  \pgfarrowsleftextend{-5\arrowsize}  
}{  
  \setlength{\arrowsize}{0.5pt}  
  \addtolength{\arrowsize}{.5\pgflinewidth}  
  \pgfpathmoveto{\pgfpoint{-5\arrowsize}{4\arrowsize}}  
  \pgfpathlineto{\pgfpointorigin}  
  \pgfpathlineto{\pgfpoint{-5\arrowsize}{-4\arrowsize}}  
  \pgfusepathqstroke  
}

\begin{document}
\begin{center}
\Large{\textbf{Diagrammatic approach to cellular automata \\ and the emergence of form with inner structure}}\\ 
~\\

\large{Vladimir Garc\'{\i}a-Morales}\\

\normalsize{}
~\\
%Institute for Advanced Study -  Technische Universit\"{a}t M\"{u}nchen,\\ Lichtenbergstr. 2a, D-85748 Garching, Germany \\

Departament de Termodin\`amica, Universitat de Val\`encia, \\ E-46100 Burjassot, Spain
\\ garmovla@uv.es
\end{center}
\small{We present a diagrammatic method to build up sophisticated cellular automata (CAs) as models of complex physical systems. The diagrams complement the mathematical approach to CA modeling, whose details are also presented here, and allow CAs in rule space to be classified according to their hierarchy of layers. Since the method is valid for any discrete operator and only depends on the alphabet size, the resulting conclusions, of general validity, apply to CAs in any dimension or order in time, arbitrary neighborhood ranges and topology. We provide several examples of the method, illustrating how it can be applied to the mathematical modeling of the emergence of order out of disorder. Specifically, we show how the the majority CA rule can be used as a building block to construct more complex cellular automata in which separate domains (with substructures having different dynamical properties) are able to emerge out of disorder and coexist in a stable manner.}
\noindent  
~\\

%\noindent complexity; predictability; time series; cellular automata; Moufang loops
\pagebreak

%\small{ \\
%%vmorales@ph.tum.de}

\section{Introduction}

Cellular automata (CAs) \cite{Wolfram,Wolfram2,Chua1,Ilachinski,Adamatzky,McIntosh,Wuensche,Toffoli,VGM1,Tokihiro1,Tokihiro6} are fully discrete dynamical systems with finite local and global phase spaces, evolving in discrete time and space. CAs provide an efficient and computationally inexpensive means of modeling complex physical systems. Although they usually constitute crude models of the physical reality, the general question arises as whether there exist increasingly sophisticated CA models able to capture both essential and subtle properties of experimental physical systems. A good deal of effort has been devoted to the identification of appropriate CA models in the huge computational space for specific needs \cite{Adamatzky}. Building on very recent work \cite{semipredo,EPL} this article advances a general and systematic method as a pathway to tackle this problem.~\\

Let a CA rule work on an alphabet of $p$ symbols, where $p\in \mathbb{N}$, $p\ge 2$ is the alphabet size. If $p$ is a composite number (i.e. not a prime number), such CA rule is $p$-decomposable \cite{semipredo}. We have presented in \cite{semipredo} a systematic way in which elementary layers can be superimposed to construct \emph{graded} CA rules where the layers are all independent. This article completes the theory in \cite{semipredo} by presenting the general construction of  \emph{non-graded} $p$-decomposable CA rules. In these CAs the layers interact in complex ways and are generally coupled yielding a wide variety of classes of models and possibilities. These couplings, however, \emph{can themselves be designed} so that the prominent features and correlations of emergent structures (that arise when iterating the models) \emph{can be predicted beforehand}. Although the output of these CA models is highly nontrivial, the systematic method presented makes them amenable to mathematical treatment and the resulting models are often easy to formulate and study. Our construction is aided by diagrams which make explicit how a given layer decomposes in sublayers and how the layers are coupled with each other. These diagrams specify universality classes of CA models that can then be automatically translated into mathematical expressions so that specific CA models belonging to these classes are constructed.  This work, thus, completes the development in \cite{semipredo}. It provides, together with that paper, a toolbox for the analysis and construction of sophisticated CA rules that can be used in the analysis of complex systems.~\\

The outline of this article is as follows. In Section \ref{CA} we present the general theory, the concept of $p$-decomposability and the diagrammatic approach that aids in the construction of models for complex systems. The relationship between the diagrams and mathematical expressions is explained. We then classify all CA rules working on alphabets where $p=p_{0}p_{1}$ has two divisors, $p_{0}$ and $p_{1}$, only. In Section \ref{examples} we give several examples where we show how CA rules can be endowed with algebraic structure, explaining how this structure is linked to the observed spatiotemporal evolution of the CA. The article is completed by giving a number of examples in which we use the majority CA rule as ground layer to build more complex CA, according to the method advanced in this article. We show how structures containing substructures of smaller size (whose dynamical behavior can be specified) arise out of disorder, thus providing a general framework to model natural processes in which form with inner structure emerges.

\section{Cellular automata: Diagrammatic approach to $p$-decomposable rules} \label{CA}

\subsection{General theory}

We first give a brief self-contained account of the main result in \cite{semipredo} that we shall need, and then proceed to describe the diagrammatic approach. In this work $S$ shall denote the set of integers in the interval $[0,p-1]$, with $p \ge 2$ being a natural number called the \emph{alphabet size}.

Let us first consider, for simplicity, a ring with $N_{s}$ sites (the following theory easily generalizes to higher dimensions and other topologies as it shall be shown in the examples). The initial condition of a CA is the vector $\mathbf{x}_{0}=(x_{0}^{0},...,x_{0}^{N_{s}-1})$, with $x^{j}_{0} \in S$, $\forall j \in [0,N_{s}-1]$ (here $j \in \mathbb{Z}$ specifies, modulo $N_{s}$, the position of a site in the ring). At each discrete time $t$, the state of the CA is given by $\mathbf{x}_{t}=(x_{t}^{0},...,x_{t}^{N_{s}-1})$ (with $x_{t}^{j} \in S$). The local spatiotemporal dynamics of the CA is a map $f:S^{l+r+1} \to S$ which provides for all $j$ the value $x_{t+1}^{j}$ as a function of some $x_{t}^{j+k}$, where $k \in [-r,l]$ specifies the position of the sites within the neighborhood of $j$. The quantity $\rho \equiv l+r+1=2\xi$ is the neighborhood range and $\xi$ is the (average) neighborhood radius. A neighborhood is called symmetric if $l=r=\xi$. The CA map $f:S^{l+r+1} \to S$ is explicitly given by the universal expression \cite{semipredo}
\begin{equation}
x_{t+1}^{j}=\mathbf{d}_{p}\left(\sum_{k=-r}^{l}p^{k+r}x_{t}^{j+k} , R \right)=\left \lfloor \frac{R}{p^{\sum_{k=-r}^{l}p^{k+r}x_{t}^{j+k}}} \right \rfloor-p \left \lfloor \frac{R}{p^{1+\sum_{k=-r}^{l}p^{k+r}x_{t}^{j+k}}} \right \rfloor
\label{themap}
\end{equation}
for each $j \in [0,N_{s}-1]$. Here, we have introduced the digit function \cite{semipredo,CHAOSOLFRAC,PHYSAFRAC} which is defined, for $p \in \mathbb{N}$, $k \in \mathbb{Z}$ and $x \in \mathbb{R}$ as  
\begin{equation}
\mathbf{d}_{p}(k,x)=\left \lfloor \frac{x}{p^{k}} \right \rfloor-p\left \lfloor \frac{x}{p^{k+1}} \right \rfloor    \label{cucuAreal}
\end{equation}
with $\lfloor \ldots \rfloor$ denoting the floor (lower closest integer) function. In Eq. (\ref{themap}) we have also introduced the Wolfram code, $R \in \mathbb{Z}$, $R\in [0, p^{p^{l+r+1}}-1]$
\begin{equation}
R \equiv \sum_{n=0}^{p^{r+l+1}-1}a_{n}p^{n}. \label{RWolf}
\end{equation} 
where the $a_{n}$'s (all $\in S$) are the coefficients which specify the CA rule. These quantities are the components of the rule vector $\mathbf{a}$. We see that $R$ is obtained from $\mathbf{a}$ through Eq. (\ref{RWolf}). Conversely, we can obtain $\mathbf{a}$ from the knowledge of $R$, since
\begin{equation}
a_{n}=\mathbf{d}_{p}(n,R) \label{coefi}
\end{equation}
i.e., the coefficients $a_{n}$ are the digits of the radix-$p$ representation of $R$.
For any $x\in \mathbb{R}$ we have
\begin{equation}
\mathbf{d}_{p}(k,p^{k}x) =\mathbf{d}_{p}(0,x)=\mathbf{d}_{p}(0,x+np) \label{pro3} 
\end{equation}
and a simple calculation shows that Eq. (\ref{themap}) is equivalent to
\begin{equation}
x_{t+1}^{j}=\sum_{n=0}^{p^{l+r+1}-1}a_{n}\mathbf{d}_{p}\left(n-\sum_{k=-r}^{l}p^{k+r}x_{t}^{j+k} , 1 \right) \label{themap2}
\end{equation}
since, at time $t$, only the term $a_{n}$ for which $n=\sum_{k=-r}^{l}p^{k+r}x_{t}^{j+k}$ contributes to the sum, all others being zero. Note that for $n, m \in \mathbb{Z}$, $\mathbf{d}_{p}\left(n-m, 1 \right)$ is a representation of the Kronecker delta $\delta_{nm}$:  We have $\delta_{nm}=1$ for $n=m$ and $\delta_{nm}=0$ for $n \ne m$.

We use the notation \cite{VGM1, VGM2}
\begin{equation}
^{l}R_{p}^{r}(x_{t}^{j}) \equiv \mathbf{d}_{p}\left(\sum_{k=-r}^{l}p^{k+r}x_{t}^{j+k} , R \right)
\end{equation}
to refer concisely to a CA rule on $p$ symbols, neighborhood radii $l$ and $r$ and Wolfram code $R$. Thus, for example $^{1}30^{1}_{2}$ is Wolfram's elementary CA rule with code $30$, and neighborhood consisting of one site to the left and one to the right of the site which is updated at the next time step. There are 256 such rules with $p=2$ and $l=r=1$.

Totalistic CA rules are a subset of the above ones and depend only on the sum over neighborhood values. In this specific case, the universal map above becomes,
\begin{equation}
x_{t+1}^{j}\equiv \ ^{l}RT_{p}^{r}(x_{t}^{j})=\mathbf{d}_{p}\left(\sum_{k=-r}^{l}x_{t}^{j+k} , R \right)
\label{themapT}
\end{equation}
or, alternatively,
\begin{equation}
x_{t+1}^{j}=\sum_{n=0}^{(p-1)(l+r+1)}a_{n}\mathbf{d}_{p}\left(n-\sum_{k=-r}^{l}x_{t}^{j+k} , 1 \right) \label{themap2T}
\end{equation}

The digit function has a decomposition property \cite{semipredo} which constitutes the main mathematical tool of this article. Let $p$ be a composite number, i.e., the product of $N$ factors $p=p_{0}p_{1}\ldots p_{N-1}=\prod_{h=0}^{N-1}p_{m}$. Let  $x_{t}^{j} \in [0,p-1]$. We have $x_{t}^{j}=\mathbf{d}_{p}(0,x_{t}^{j})$. Then,
\begin{eqnarray}
\mathbf{d}_{p}(0,x_{t}^{j})&=&x_{t}^{j}-p\left \lfloor \frac{x_{t}^{j}}{p} \right \rfloor=x_{t}^{j}-p_{0}p_{1}\ldots p_{N-1}\left \lfloor \frac{x_{t}^{j}}{p_{0}p_{1}\ldots p_{N-1}} \right \rfloor
 \nonumber  \\
&=&x_{t}^{j}-p_{0}\left \lfloor \frac{x_{t}^{j}}{p_{0}} \right \rfloor+p_{0}\left \lfloor \frac{x_{t}^{j}}{p_{0}} \right \rfloor-p_{0}p_{1}\left \lfloor \frac{x_{t}^{j}}{p_{0}p_{1}} \right \rfloor+p_{0}p_{1}\left \lfloor \frac{x_{t}^{j}}{p_{0}p_{1}} \right \rfloor-\ldots \nonumber \\
&&-p_{0}p_{1}\ldots p_{N-1}\left \lfloor \frac{x_{t}^{j}}{p_{0}p_{1}\ldots p_{N-1}} \right \rfloor
 \nonumber \\
&=&\mathbf{d}_{p_{0}}\left(0, x_{t}^{j} \right)+p_{0}\mathbf{d}_{p_{1}}\left(0, \frac{x_{t}^{j}}{p_{0}} \right)+\ldots+p_{0}p_{1}\ldots p_{N-2}\mathbf{d}_{p_{N-1}}\left(0, \frac{x_{t}^{j}}{p_{0}p_{1}\ldots p_{N-2}} \right) \nonumber \\
&=&
\sum_{h=0}^{N-1}\mathbf{d}_{p_{h}}\left(0, \frac{x_{t}^{j}}{\prod_{m=0}^{h-1}p_{m}} \right)\prod_{m=0}^{h-1}p_{m}    
\end{eqnarray}
Hence, by defining local integer variables that take values in the interval $[0,p_{h}-1]$
\begin{equation}
y_{t}^{(h),j}\equiv \mathbf{d}_{p_{h}}\left(0, \frac{x_{t}^{j}}{\prod_{m=0}^{h-1}p_{m}} \right) \qquad h=0,1,\ldots, N-1 \label{ini0}
\end{equation}
we can write \cite{semipredo}
\begin{equation}
x_{t}^{j}=y_{t}^{(0),j}+p_{0}y_{t}^{(1),j}+p_{0}p_{1}y_{t}^{(2),j}+\ldots + p_{0}p_{1}\ldots p_{N-2}y_{t}^{(N-1),j}=\sum_{h=0}^{N-1}y_{t}^{(h),j}\prod_{m=0}^{h-1}p_{m}
\end{equation}
This expression coincides with Cantor's mixed-radix representation of the integer number $x_{t}^{j}$ in terms of the factors $p_{h}$ \cite{Cantor69}. Of course, at time $t+1$ we have, similarly,
\begin{equation}
x_{t+1}^{j}=\sum_{h=0}^{N-1}y_{t+1}^{(h),j}\prod_{m=0}^{h-1}p_{m} \label{atp1gen}
\end{equation}
and, hence, the transformation $x_{t}^{j}\to x_{t+1}^{j}$ on $p$ symbols is equivalent to the transformation $(y^{(0),j}_{t},\ldots, y^{(h),j}_{t},\ldots, y^{(N-1),j}_{t}) \to (y^{(0),j}_{t+1},\ldots, y^{(h),j}_{t+1},\ldots, y^{(N-1),j}_{t+1})$ on $N$-tuples of variables $y^{(h),j}_{t}\in [0,p_{h}-1]$. From Eqs. (\ref{themap}), (\ref{ini0}) and (\ref{atp1gen}), we have 
\begin{eqnarray}
y^{(h),j}_{t+1} &=& \mathbf{d}_{p_{h}}\left(0, \frac{1}{\prod_{m=0}^{h-1}p_{m}}x_{t+1}^{j} \right) \label{couply0} \\
&=&\mathbf{d}_{p_{h}}\left(0, \frac{1}{\prod_{m=0}^{h-1}p_{m}}\mathbf{d}_{p}\left(\sum_{k=-r}^{l}p^{k+r}\left[\sum_{s=0}^{N-1}y_{t}^{(s),j+k}\prod_{m=0}^{s-1}p_{m}\right] ,\ R \right) \right) \label{couply}
\end{eqnarray}
which we call the \emph{h-layer value} of the CA $^{l}R_{p}^{r}(x_{t}^{j})$.

We note that the above decomposition is independent of the topology and dimensionality of the model CA and is only based on the alphabet structure. If $y_{t+1}^{(h),j}$ on layer $h$ only depends at time $t+1$ on the $y_{t}^{(h),j}$  variables of that same layer $h$, Eq. (\ref{couply}) reduces to
\begin{eqnarray}
y_{t+1}^{(h),j}&=&\mathbf{d}_{p_{h}}\left(0, \frac{1}{\prod_{m=0}^{h-1}p_{m}}\mathbf{d}_{p}\left(\sum_{k=-r}^{l}p^{k+r}y_{t}^{(h),j+k},\ R \right) \right)  \label{layerCA} \\
&=&\mathbf{d}_{p_{h}}\left(\sum_{k=-r}^{l}p_{h}^{k+r}y_{t}^{(h),j+k},\ A_{p_{h}}^{(h)} \right)=
\mathbf{d}_{p_{h}}\left(\sum_{k=-r}^{l}p_{h}^{k+r}\mathbf{d}_{p_{h}}\left(0, \frac{x_{t}^{j}}{\prod_{m=0}^{h-1}p_{m}} \right),\ A_{p_{h}}^{(h)} \right) \nonumber
\end{eqnarray}
i.e. each layer value behaves as a CA with Wolfram code $A_{p_{h}}^{(h)}$ that is independent of the other layers. We call them \emph{layer CAs}. If all layers are layer CAs the CA rule $^{l}R_{p}^{r}$ is called a \emph{graded rule} \cite{semipredo}. In general, however, the layer values are nonlinearly coupled as seen in Eq. (\ref{couply}). In this general case we say that $^{l}R_{p}^{r}$ is a $p$-decomposable rule. All graded CA rules are $p$-decomposable rules but the contrary is not true.

We observe that any factor $p_{h}$ of $p$ can also be composite. In that case, any layer $A^{(h)}_{p_{h}}$ can be further decomposed in sublayers following the above scheme. We thus introduce the pair of integers $(g,h)$ called the \emph{rank} $g$ and the \emph{layer index} $h$. We order the layers in ranks. If the rank of a layer is $g$, the rank of a sublayer is $g+1$.  The  rank $g=0$ is the CA model of the system. On each rank $g$, the layer with $h=0$ is called the \emph{ground layer} and any layer with $h>0$ a \emph{lifted layer}. If the layers are CA then we speak about a ground CA layer and a lifted CA layer respectively. 

The following are the rules to construct diagrams: 
\begin{itemize}
\item At the top of the diagram (rank $g=0$) we put the CA rule $^{l}R^{r}_{p}$ that constitutes the \emph{model}.
\item Layers with increasing rank are arranged vertically from top to bottom.
\item Layers with increasing index $h$ are arranged from left to right. Thus, the ground layer is the leftmost node of a certain rank.
\item We label any layer CA with a solid node and an arbitrary layer that is not a CA with a hollow node.
\item Undirected edges are vertically drawn to relate layers of rank $g$ to their decompositions into layers of rank $g+1$. Such edges are to be read as ``layer with rank $g$ decomposes as layers $A, B, C, \ldots$ with rank $g+1$''.
\item We omit the radii ($l$ and $r$) of any layers different to the model (at the top) of the diagram. If the radii of the other nodes are layer-dependent, i.e., if one has $l^{(g,h)}$, $r^{(g,h)}$, for the radii, then the model has $l=\max l^{(g,h)}$ and $r=\max r^{(g,h)}$. For each layer, only the 'coordinates' $(g,h)$ and the alphabet size $p_{h}$ are indicated.
\item Directed edges indicate the dependence of a layer on other layers. An outgoing arrow from a node $A$ to a node $B$ means that $A$ \emph{influences} $B$ but is independent of $B$. The incoming arrow at $B$ means that the value of $B$ depends on the value (or configuration) at $A$.  
\end{itemize}

\begin{figure}
\includegraphics[width=1.0 \textwidth]{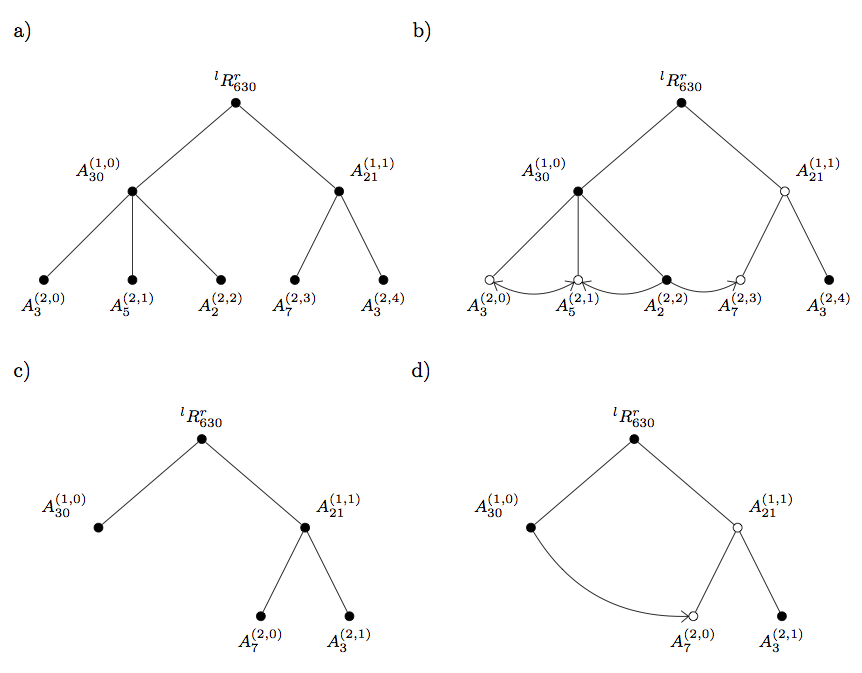}
\caption{\scriptsize{a) A fully decomposed diagram of a graded CA rule with $p=630$. b) A possible $p$-decomposable non-graded CA rule derived from a). In c) and d) two partially decomposed diagrams compatible with a) and b), respectively, are shown.
}} \label{arbitrary}
\end{figure}

In Fig. \ref{arbitrary} a) we see the full decomposition of a graded CA rule on $p=630$ symbols. It contains only undirected edges and layer CAs. The rule has two nonzero ranks, $g=1$ with a ground layer CA on $p_{0}=30$ symbols and a lifted layer CA on $p_{1}=21$ symbols. The former layer is further decomposed into three rank-2 layer CAs,  $A^{(2,0)}_{3}$, $A^{(2,1)}_{5}$, $A^{(2,2)}_{2}$. Layer $A^{(1,1)}_{21}$ is decomposed in two rank-2 layer CAs. We note that we could have stopped the decomposition at any of the nodes with rank $g=1$. If the decomposition is shown up to the non-decomposable sublayers (over a prime number of symbols) we say that the CA diagram is \emph{fully decomposed}. There is a unique fully decomposed diagram for each given CA model. If some layers over composite $p_{h}$ are not further decomposed, we say that the CA diagram is \emph{partially decomposed}. Any fully decomposed diagram can be thus  embedded in a partially decomposed diagram and the former are a subset of the latter. Often, only partial decompositions are relevant. In Fig. \ref{arbitrary} c) a partially decomposed diagram that contains the one of Fig. \ref{arbitrary} a) is plotted.

In Fig. \ref{arbitrary} b) a $p$-decomposable non-graded CA rule is shown. We see that, although the number of nodes and the alphabets of the layers are the same as in Fig. \ref{arbitrary} a), not all layers are layer CAs, and some layers depend on others: We see that layer CA $A^{(2,2)}_{2}$ influences two of the layers, while layers  $A^{(2,0)}_{3}$ and $A^{(2,1)}_{5}$ are interdependent. In Fig. \ref{arbitrary} d) a partially decomposed diagram that contains the one of Fig. \ref{arbitrary} b) as a specific case, is plotted.

We summarize some observations:
\begin{itemize}
\item The top rule in a diagram is the CA model. The inputs of this model are $x_{t}^{j-r}, \ldots, x_{t}^{j}$, $\ldots, x_{t}^{j+l}$ and the output $x_{t+1}^{j}$ at every $j$ and $t$. For the top layer, we always write $x_{t}^{j}$ and $x_{t+1}^{j}$ instead of $y_{t}^{(0,0),j}$, $y_{t+1}^{(0,0),j}$, respectively.
\item A graded CA rule contains only layer CAs and undirected edges.
\item The product of the $p_{h}$'s of rank-$(g+1)$ layers coming from a single node at rank $g$ equals $p$ of that node.
\item From a layer CA only outgoing arrows or undirected edges are possible.
\item From a layer that is not a CA, undirected edges and incoming and outgoing arrows are possible. If it has only undirected edges then it must exhibit an incoming arrow in at least one partial decomposition of the tree.
\item If a node corresponds to a layer on a prime number of symbols it cannot be further decomposed.
\end{itemize}

\begin{figure}
\includegraphics[width=0.8 \textwidth]{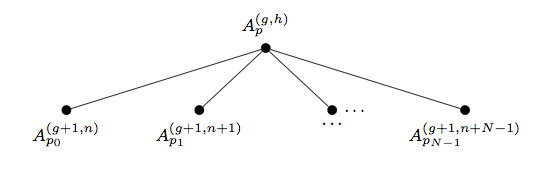}
\caption{\scriptsize{A diagram for an arbitrary graded rule $A_{p}^{(g,h)}$ at node $(g,h)$  of a tree. 
}} \label{gsgp1}
\end{figure}

Since graded rules constitute the backbone of any tree, it is useful first to see how a diagram representing an arbitrary graded rule at a node $(g,h)$ can be translated to a mathematical model. Such arbitrary building block from which to construct trees as the one in Figs. \ref{arbitrary} is shown in Fig. \ref{gsgp1}. Let us assume that a node with coordinates $(g,h)$ in a diagram splits as shown in Fig. \ref{gsgp1}. From Eqs. (\ref{ini0}) to (\ref{layerCA}) we have
\begin{equation}
y_{t+1}^{(g,h),j}=\sum_{s=0}^{N-1}y_{t+1}^{(g+1,n+s),j}\prod_{m=0}^{s-1}p_{m} \label{atp1genB}
\end{equation}
where
\begin{equation}
y_{t+1}^{(g+1,n+s),j}=\mathbf{d}_{p_{s}}\left(\sum_{k=-r}^{l}p_{s}^{k+r}y_{t}^{(g+1,n+s),j+k},\ A_{p_{s}}^{(g+1,n+s)} \right) \qquad s=0,1,\ldots, N-1 \label{couplyB}
\end{equation}
and
\begin{equation}
y_{t}^{(g+1,n+s),j}\equiv \mathbf{d}_{p_{s}}\left(0, \frac{y_{t}^{(g,h),j}}{\prod_{m=0}^{s-1}p_{m}} \right) \qquad s=0,1,\ldots, N-1 \label{ini0B}
\end{equation}

These expressions are thus the mathematical transposition of the tree in Fig. \ref{gsgp1} and are to be interpreted as follows. The value $y_{t+1}^{(g,h),j}$ on the node governed by $A_{p}^{(g,h)}$ is known from the previous value $y_{t}^{(g,h),j}$, by replacing the latter in Eq. (\ref{ini0B}) to obtain all the $y_{t}^{(g+1,n+s),j}$'s on the rank $g+1$ at time $t$. Then, all these quantities are replaced in Eq. (\ref{couplyB}) to obtain all $N$ quantities $y_{t+1}^{(g+1,n+s),j}$ at the next time step and, from them, the value of $y_{t+1}^{(g,h),j}$ is finally obtained from Eq. (\ref{atp1genB}). Because of the independence of the layer CAs within the CA model, we say that the latter is the direct sum of its layer CAs and write
\begin{equation}
A_{p}^{(g,h)}=A_{p_{0}}^{(g+1,n)}\oplus A_{p_{1}}^{(g+1,n+1)}\oplus \ldots \oplus A_{p_{0}}^{(g+1,n+N-1)}
\end{equation}

The couplings between layers are represented through arrows as described above. Let an arrow start from a node with coordinates $(g',h')$ and end in a node with coordinates $(g,h)$. 
Thus, layer $h$ is influenced by layer $h'$. The mathematical general way to express this relationship is obtained, by using Eq. (\ref{themap2}), as
\begin{equation}
y_{t+1}^{(g,h),j}=\sum_{n'=0}^{p_{h'}^{l+r+1}-1} \mathbf{d}_{p_{h}}\left(\sum_{k=-r}^{l}p_{h}^{k+r}y_{t}^{(g,h),j+k} , A_{n', p_h}^{(g,h)} \right)
\mathbf{d}_{p_{h'}}\left(n'-\sum_{k=-r}^{l}p_{h'}^{k+r}y_{t}^{(g',h'),j+k} , 1 \right) \label{themapcoup}
\end{equation}
where the $A_{n', p_h}^{(g',h')}$ are integers in the interval $[0,p_{h}^{p_{h}^{l+r+1}}-1]$. If all  $A_{n', p_h}^{(g',h')}$ are independent of $n'$, then, the layer $(g',h')$ becomes a layer CA with Wolfram code $A_{n', p_h}^{(g,h)}=A_{p_{h}}^{(g,h)}$. In this case, from this latter expression, since $\sum_{n'=0}^{p_{h'}^{l+r+1}-1} \mathbf{d}_{p_{h'}}\left(n'-\sum_{k=-r}^{l}p_{h'}^{k+r}y_{t}^{(g',h'),j+k} , 1 \right)=1$, we have
\begin{equation}
y_{t+1}^{(g',h'),j}=\mathbf{d}_{p_{h'}}\left(\sum_{k=-r}^{l}p_{h'}^{k+r}y_{t}^{(g',h'),j+k} , A_{p_{h}}^{(g,h)} \right) \label{themapcoupin}
\end{equation} 
Eq. (\ref{themapcoup}) thus describes any hollow node in the tree and Eq. (\ref{themapcoupin}) a solid node (as we have seen above).
If there are more incoming arrows at node $(g,h)$ coming from nodes e.g. at $(g',h')$ and $(g'', h'')$, Eq. (\ref{themapcoup}) generalizes as
\begin{eqnarray}
y_{t+1}^{(g,h),j}&=&\sum_{n''=0}^{p_{h''}^{l+r+1}-1}\sum_{n'=0}^{p_{h'}^{l+r+1}-1} \mathbf{d}_{p_{h}}\left(\sum_{k=-r}^{l}p_{h}^{k+r}y_{t}^{(g,h),j+k} , A_{n'n'', p_h}^{(g,h)} \right)\times \label{themapcoup2} \\
&&\times \mathbf{d}_{p_{h'}}\left(n'-\sum_{k=-r}^{l}p_{h'}^{k+r}y_{t}^{(g',h'),j+k} , 1 \right)\mathbf{d}_{p_{h''}}\left(n''-\sum_{k=-r}^{l}p_{h''}^{k+r}y_{t}^{(g'',h''),j+k} , 1 \right) \nonumber
\end{eqnarray}
where, again, the $A_{n'n'', p_h}^{(g',h')}$ are integers in the interval $[0,p_{h}^{p_{h}^{l+r+1}}-1]$. The above expression generalizes to an arbitrary number of incoming arrows.

A remark is in order on how to calculate the integer numbers $A_{n', p_h}^{(g,h)}$ from an arbitrary function $f_{n'}: S^{l+r+1} \to S$, $0\le f_{n'} \le p-1$ ($f_{n'} \in \mathbb{Z}$)
\begin{equation}
f_{n'}\left(y_{t}^{(g,h),j-r},y_{t}^{(g,h),j-r+1},\ldots, y_{t}^{(g,h),j-r}, \ldots, y_{t}^{(g,h),j+l-1}, y_{t}^{(g,h),j+l}\right) 
\end{equation}
Let $n=\sum_{k=-r}^{l}p_{h}^{k+r}y_{t}^{(g,h),j+k}$. We have $y_{t}^{(g,h),j+k}=\mathbf{d}_{p_{h}}(k+r,n)$. Therefore,
\begin{equation}
A_{n', p_h}^{(g,h)}=\sum_{n=0}^{p_{h}^{l+r+1}-1}p_{h}^{n}f_{n'}\left(\mathbf{d}_{p_{h}}(0,n),
\ldots, \mathbf{d}_{p_{h}}(r,n), \ldots, \mathbf{d}_{p_{h}}(l+r,n)\right) 
\end{equation}
since
\begin{equation}
\mathbf{d}_{p_{h}}\left(\sum_{k=-r}^{l}p_{h}^{k+r}y_{t}^{(g,h),j+k} , A_{n', p_h}^{(g,h)} \right)=\mathbf{d}_{p_{h}}\left(n, A_{n', p_h}^{(g,h)} \right)=f_{n'}
\end{equation}

Having explained how to pass from diagrams to mathematical expressions and vice versa,
this abstract theory is now completed. The CA modeling through the diagrammatic approach proceeds through the following steps
\begin{itemize}
\item 1. Determine how many layers are needed and construct the diagram of a graded CA rule with all those layers. Transpose the diagram to a mathematical model employing Eqs. (\ref{atp1genB}) to (\ref{ini0B}) as building blocks for each branch in the diagram.
\item 2. Draw the arrows needed to couple the corresponding layers. For each layer $(g',h')$ with an incoming arrow, replace the corresponding mathematical expression obtained from the graded rule for $y_{t+1}^{(g',h'),j}$ by Eq. (\ref{themapcoupin}).
\end{itemize}

Let us illustrate this approach with reference to the trees in Fig. \ref{arbitrary}. 
The tree in Fig. \ref{arbitrary} a) is a graded CA rule and we can write out the mathematical model as follows
\begin{eqnarray}
x_{t+1}^{j}&=&y_{t+1}^{(1,0),j}+30y_{t+1}^{(1,1),j} \label{ar0} \\
y_{t+1}^{(1,0),j}&=& y_{t+1}^{(2,0),j}+3y_{t+1}^{(2,1),j}+15y_{t+1}^{(2,2),j} \label{ar10} \\
y_{t+1}^{(1,1),j}&=& y_{t+1}^{(2,3),j}+7y_{t+1}^{(2,4),j} \label{ar10} \\
y_{t+1}^{(2,0),j}&=& \mathbf{d}_{3}\left(\sum_{k=-r}^{l}3^{k+r}y_{t}^{(2,0),j+k},\ A_{3}^{(2,0)} \right)  \qquad \qquad   \\
y_{t+1}^{(2,1),j}&=& \mathbf{d}_{5}\left(\sum_{k=-r}^{l}5^{k+r}y_{t}^{(2,1),j+k},\ A_{5}^{(2,1)} \right)  \qquad \qquad   \\
y_{t+1}^{(2,2),j}&=& \mathbf{d}_{2}\left(\sum_{k=-r}^{l}2^{k+r}y_{t}^{(2,2),j+k},\ A_{2}^{(2,2)} \right)  \qquad \qquad    \\
y_{t+1}^{(2,3),j}&=&  \mathbf{d}_{7}\left(\sum_{k=-r}^{l}7^{k+r}y_{t}^{(2,3),j+k},\ A_{7}^{(2,3)} \right)  \qquad \qquad  \\
y_{t+1}^{(2,4),j}&=&  \mathbf{d}_{3}\left(\sum_{k=-r}^{l}3^{k+r}y_{t}^{(2,4),j+k},\ A_{3}^{(2,4)} \right) \qquad \qquad  
\end{eqnarray}
\begin{eqnarray}
y_{t}^{(2,0),j} &=& \mathbf{d}_{3}\left(0, y_{t}^{(1,0),j} \right) \qquad y_{t}^{(2,1),j} = \mathbf{d}_{5}\left(0, \frac{y_{t}^{(1,0),j}}{3} \right) \qquad y_{t}^{(2,2),j} = \mathbf{d}_{2}\left(0, \frac{y_{t}^{(1,0),j}}{15} \right) \nonumber \\&& \\
y_{t}^{(2,3),j} &=& \mathbf{d}_{7}\left(0, y_{t}^{(1,1),j} \right) \qquad y_{t}^{(2,4),j} = \mathbf{d}_{3}\left(0, \frac{y_{t}^{(1,1),j}}{7} \right)  \\
y_{t}^{(1,0),j} &=& \mathbf{d}_{30}\left(0, x_{t}^{j} \right) \qquad \quad \ y_{t}^{(1,1),j} = \mathbf{d}_{21}\left(0, \frac{x_{t}^{j}}{30} \right)  \label{thelastone}
\end{eqnarray}
Thus, computations with such a model are carried from bottom to top, starting from Eqs. (\ref{thelastone}) for an input value of $x_{t}^{j} \in [0,629]$, $\forall j \in [0,N_{s}-1]$ and the five integers $A_{3}^{(2,0)} \in [0,3^{3^{l+r+1}}-1]$, $A_{5}^{(2,1)} \in [0,5^{5^{l+r+1}}-1]$, $A_{2}^{(2,2)} \in [0,2^{2^{l+r+1}}-1]$, $A_{7}^{(2,3)} \in [0,7^{7^{l+r+1}}-1]$ and $A_{3}^{(2,4)} \in [0,3^{3^{l+r+1}}-1]$. Thus all all quantities are calculated from bottom to top so that, at the end $x_{t+1}^{j} \in [0,629]$ is obtained from Eq. (\ref{ar0}). This system of equations is the mathematical equivalent to the diagram in Fig. \ref{arbitrary} a).  

It is now straightforward to construct the family of CA rules described by the tree in Fig. \ref{arbitrary} b). We have just only to provide expressions for $y_{t+1}^{(g,h),j}$ for all those layers that have an incoming arrow. In the tree in Fig. \ref{arbitrary} b) there are three such layers, whose values at time $t+1$ are given by $y_{t+1}^{(2,0),j}$, $y_{t+1}^{(2,1),j}$ and $y_{t+1}^{(2,3),j}$. Thus, by using Eqs. (\ref{themapcoup}) and (\ref{themapcoup2}) 
\begin{eqnarray}
y_{t+1}^{(2,0),j}&=&\sum_{n'=0}^{5^{l+r+1}-1} \mathbf{d}_{3}\left(\sum_{k=-r}^{l}3^{k+r}y_{t}^{(2,0),j+k} , A_{n', p_h}^{(2,0)} \right)
\mathbf{d}_{5}\left(n'-\sum_{k=-r}^{l}5^{k+r}y_{t}^{(2,1),j+k} , 1 \right)  \nonumber \\
y_{t+1}^{(2,1),j}&=&\sum_{n''=0}^{2^{l+r+1}-1}\sum_{n'=0}^{3^{l+r+1}-1} \mathbf{d}_{5}\left(\sum_{k=-r}^{l}5^{k+r}y_{t}^{(2,1),j+k} , A_{n'n'', p_h}^{(2,1)} \right)\times  \\
&&\times \mathbf{d}_{3}\left(n'-\sum_{k=-r}^{l}3^{k+r}y_{t}^{(2,0),j+k} , 1 \right)\mathbf{d}_{2}\left(n''-\sum_{k=-r}^{l}2^{k+r}y_{t}^{(2,2),j+k} , 1 \right) \nonumber \\
y_{t+1}^{(2,3),j}&=&\sum_{n'=0}^{2^{l+r+1}-1} \mathbf{d}_{7}\left(\sum_{k=-r}^{l}7^{k+r}y_{t}^{(2,3),j+k} , A_{n', p_h}^{(2,3)} \right)
\mathbf{d}_{2}\left(n'-\sum_{k=-r}^{l}2^{k+r}y_{t}^{(2,2),j+k} , 1 \right)  \nonumber
\end{eqnarray}
and these expressions replace the corresponding ones within the graded rule above.
In this way we account for all possible couplings among layers described by the tree in Fig. \ref{arbitrary} b). Note that all other equations remain unchanged, as in the graded rule.

Models of partially decomposed diagrams are considerably simplified compared to the ones corresponding to fully developed diagrams. Therefore, the specific choice to which extent a diagram is to be developed depends on the specific problem at hand. The diagram in Fig. \ref{arbitrary} c) is mathematically described by the following system of equations
\begin{eqnarray}
x_{t+1}^{j}&=&y_{t+1}^{(1,0),j}+30y_{t+1}^{(1,1),j} \label{ar0B} \\
y_{t+1}^{(1,1),j}&=& y_{t+1}^{(2,0),j}+7y_{t+1}^{(2,1),j} \label{ar10B} \\
y_{t+1}^{(2,0),j}&=&  \mathbf{d}_{7}\left(\sum_{k=-r}^{l}7^{k+r}y_{t}^{(2,0),j+k},\ A_{7}^{(2,0)} \right)  \qquad \qquad  \\
y_{t+1}^{(2,1),j}&=&  \mathbf{d}_{3}\left(\sum_{k=-r}^{l}3^{k+r}y_{t}^{(2,1),j+k},\ A_{3}^{(2,1)} \right) \qquad \qquad  \\
y_{t+1}^{(1,0),j}&=&  \mathbf{d}_{30}\left(\sum_{k=-r}^{l}30^{k+r}y_{t}^{(1,0),j+k},\ A_{30}^{(1,0)} \right) \qquad \qquad  \\
y_{t}^{(2,0),j} &=& \mathbf{d}_{7}\left(0, y_{t}^{(1,1),j} \right) \qquad y_{t}^{(2,1),j} = \mathbf{d}_{3}\left(0, \frac{y_{t}^{(1,1),j}}{7} \right)  \\
y_{t}^{(1,0),j} &=& \mathbf{d}_{30}\left(0, x_{t}^{j} \right) \qquad \quad \ y_{t}^{(1,1),j} = \mathbf{d}_{21}\left(0, \frac{x_{t}^{j}}{30} \right)  \label{thelastoneB}
\end{eqnarray}
Finally, the diagram in Fig. \ref{arbitrary} d) corresponds to the following mathematical model
\begin{eqnarray}
x_{t+1}^{j}&=&y_{t+1}^{(1,0),j}+30y_{t+1}^{(1,1),j} \label{ar0C} \\
y_{t+1}^{(1,1),j}&=& y_{t+1}^{(2,0),j}+7y_{t+1}^{(2,1),j} \label{ar10C} \\
y_{t+1}^{(2,0),j}&=& \sum_{n'=0}^{30^{l+r+1}-1} \mathbf{d}_{7}\left(\sum_{k=-r}^{l}7^{k+r}y_{t}^{(2,0),j+k} , A_{n', p_h}^{(2,0)} \right)
\mathbf{d}_{30}\left(n'-\sum_{k=-r}^{l}30^{k+r}y_{t}^{(1,0),j+k} , 1 \right)  \nonumber \\ && \\
y_{t+1}^{(2,1),j}&=&  \mathbf{d}_{3}\left(\sum_{k=-r}^{l}3^{k+r}y_{t}^{(2,1),j+k},\ A_{3}^{(2,1)} \right)  \qquad \qquad  \\
y_{t+1}^{(1,0),j}&=&  \mathbf{d}_{30}\left(\sum_{k=-r}^{l}30^{k+r}y_{t}^{(1,0),j+k},\ A_{30}^{(1,0)} \right) \qquad \qquad  
\end{eqnarray}
\begin{eqnarray}
y_{t}^{(2,0),j} &=& \mathbf{d}_{7}\left(0, y_{t}^{(1,1),j} \right) \qquad y_{t}^{(2,1),j} = \mathbf{d}_{3}\left(0, \frac{y_{t}^{(1,1),j}}{7} \right)  \\
y_{t}^{(1,0),j} &=& \mathbf{d}_{30}\left(0, x_{t}^{j} \right) \qquad \quad \ y_{t}^{(1,1),j} = \mathbf{d}_{21}\left(0, \frac{x_{t}^{j}}{30} \right)  \label{thelastoneC}
\end{eqnarray}

In this way we obtain simple constructions of CA models on 630 symbols. We devote the rest of the article to examples to illustrate the above approach. These are presented in the next section.

\subsection{Classification of CA rules for $p$ having a small number of proper divisors} 

All CA rules can be classified according to the above approach and we carry out now this in the simplest cases for $p$ having a small number of divisors. It is clear, from all above, that the `tree' of a non-decomposable rule, i.e. a rule where the alphabet size $p$ is a prime number, has the trivial representation
\begin{center}
\includegraphics[width=0.05 \textwidth]{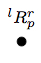}
\end{center}
The mathematical expression for these CA is just Eq. (\ref{themap}) because $p$ has not proper divisors and, hence, no layers other than the trivial one (i.e. the CA). All Boolean CA ($p=2$) fall in this category. 

\begin{figure} 
\includegraphics[width=0.9 \textwidth]{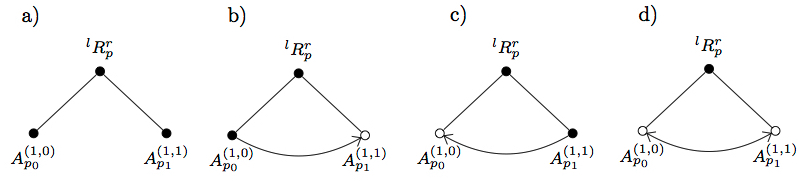}
\caption{\scriptsize{Classification of all $^{l}R^{r}_{p}$ CA rules on $p=p_{0}p_{1}$ symbols. a) Graded rules, b) to d) $p$-decomposable non-graded CA rules.
}} \label{2times2}
\end{figure}

We now consider decomposable CA rules $^{l}R^{r}_{p}$ on $p=p_{0}p_{1}$ symbols, with $p_{0}$ and $p_{1}$ both prime (not necessarily distinct). There are $p^{p^{l+r+1}}=(p_{0}p_{1})^{(p_{0}p_{1})^{l+r+1}}$ one-dimensional deterministic CA rules of this kind 
and they can all be classified according to the diagrams shown in Fig. \ref{2times2}.
The CA rules with tree as in a) are graded CA rules. All others are $p$-decomposable non graded CA rules.

The graded rules, diagram in Fig. \ref{2times2} a), have all the following mathematical model
\begin{eqnarray}
x_{t+1}^{j}&=&y_{t+1}^{(1,0),j}+p_{0}y_{t+1}^{(1,1),j} \label{2t2a0}\\
y_{t+1}^{(1,0),j}&=& \mathbf{d}_{p_{0}}\left(\sum_{k=-r}^{l}p_{0}^{k+r}y_{t}^{(1,0),j+k},\ A_{p_{0}}^{(1,0)} \right)     \label{2t2a1} \\
y_{t+1}^{(1,1),j}&=& \mathbf{d}_{p_{1}}\left(\sum_{k=-r}^{l}p_{1}^{k+r}y_{t}^{(1,1),j+k},\ A_{p_{1}}^{(1,1)} \right)    \label{2t2a2} \\
y_{t}^{(1,0),j} &=& \mathbf{d}_{p_{0}}\left(0, x_{t}^{j} \right) \qquad \quad y_{t}^{(1,1),j} = \mathbf{d}_{p_{1}}\left(0, \frac{x_{t}^{j}}{p_{0}} \right)  \label{2t2a3}
\end{eqnarray}

These equations can be more concisely expressed in a single equation as
\begin{equation}
x_{t+1}^{j}=\mathbf{d}_{p_{0}}\left(\sum_{k=-r}^{l}p_{0}^{k+r}\mathbf{d}_{p_{0}}\left(0, x_{t}^{j+k} \right),\ A_{p_{0}}^{(1,0)} \right)+p_{0}\mathbf{d}_{p_{1}}\left(\sum_{k=-r}^{l}p_{1}^{k+r}\mathbf{d}_{p_{1}}\left(0, \frac{x_{t}^{j+k}}{p_{0}} \right),\ A_{p_{1}}^{(1,1)} \right) \label{p0p1gra}
\end{equation}
which is equivalent to the above system. 
 
The CAs described by Fig. \ref{2times2} b) have the following general form
\begin{eqnarray}
x_{t+1}^{j}&=&y_{t+1}^{(1,0),j}+p_{0}y_{t+1}^{(1,1),j} \label{2t2b0}\\
y_{t+1}^{(1,0),j}&=& \mathbf{d}_{p_{0}}\left(\sum_{k=-r}^{l}p_{0}^{k+r}y_{t}^{(1,0),j+k},\ A_{p_{0}}^{(1,0)} \right)   \label{2t2b1} \\
y_{t+1}^{(1,1),j}&=& \sum_{n'=0}^{p_{0}^{l+r+1}-1} \mathbf{d}_{p_{1}}\left(\sum_{k=-r}^{l}p_{1}^{k+r}y_{t}^{(1,1),j+k} , A_{n', p_1}^{(1,1)} \right)
\mathbf{d}_{p_{0}}\left(n'-\sum_{k=-r}^{l}p_{0}^{k+r}y_{t}^{(1,0),j+k} , 1 \right) \nonumber \\ && \\
y_{t}^{(1,0),j} &=& \mathbf{d}_{p_{0}}\left(0, x_{t}^{j} \right) \qquad \qquad y_{t}^{(1,1),j} = \mathbf{d}_{p_{1}}\left(0, \frac{x_{t}^{j}}{p_{0}} \right) \label{2t2b2}
\end{eqnarray}
and it is equally straightforward to write the general form corresponding to the diagram in Fig. \ref{2times2} c) 
\begin{eqnarray}
x_{t+1}^{j}&=&y_{t+1}^{(1,0),j}+p_{0}y_{t+1}^{(1,1),j} \label{2t2c0}\\
y_{t+1}^{(1,0),j}&=& \sum_{n'=0}^{p_{1}^{l+r+1}-1} \mathbf{d}_{p_{0}}\left(\sum_{k=-r}^{l}p_{0}^{k+r}y_{t}^{(1,0),j+k} , A_{n', p_0}^{(1,0)} \right)
\mathbf{d}_{p_{1}}\left(n'-\sum_{k=-r}^{l}p_{1}^{k+r}y_{t}^{(1,1),j+k} , 1 \right) \label{2t2c1} \nonumber \\ &&\\
y_{t+1}^{(1,1),j}&=& \mathbf{d}_{p_{1}}\left(\sum_{k=-r}^{l}p_{1}^{k+r}y_{t}^{(1,1),j+k},\ A_{p_{1}}^{(1,1)} \right)   \label{2t2c2} \\
y_{t}^{(1,0),j} &=& \mathbf{d}_{p_{0}}\left(0, x_{t}^{j} \right) \qquad \qquad y_{t}^{(1,1),j} = \mathbf{d}_{p_{1}}\left(0, \frac{x_{t}^{j}}{p_{0}} \right) \label{2t2c3}
\end{eqnarray}
The Moufang loop example  with $p=12$ in \cite{semipredo} belongs to this class (if one constructs a partially developed tree with $p_{0}=6$ and $p_{1}=2$). Finally Fig. \ref{2times2} d) is mathematically transposed as
\begin{eqnarray}
x_{t+1}^{j}&=&y_{t+1}^{(1,0),j}+p_{0}y_{t+1}^{(1,1),j} \label{2t2d0}\\
y_{t+1}^{(1,0),j}&=& \sum_{n'=0}^{p_{1}^{l+r+1}-1} \mathbf{d}_{p_{0}}\left(\sum_{k=-r}^{l}p_{0}^{k+r}y_{t}^{(1,0),j+k} , A_{n', p_0}^{(1,0)} \right)
\mathbf{d}_{p_{1}}\left(n'-\sum_{k=-r}^{l}p_{1}^{k+r}y_{t}^{(1,1),j+k} , 1 \right) \nonumber\\ && \label{2t2d1} \\
y_{t+1}^{(1,1),j}&=& \sum_{n''=0}^{p_{0}^{l+r+1}-1} \mathbf{d}_{p_{1}}\left(\sum_{k=-r}^{l}p_{1}^{k+r}y_{t}^{(1,1),j+k} , A_{n'', p_1}^{(1,1)} \right)
\mathbf{d}_{p_{0}}\left(n''-\sum_{k=-r}^{l}p_{0}^{k+r}y_{t}^{(1,0),j+k} , 1 \right) \nonumber\\ &&  \label{2t2d2} \\
y_{t}^{(1,0),j} &=& \mathbf{d}_{p_{0}}\left(0, x_{t}^{j} \right) \qquad \qquad y_{t}^{(1,1),j} = \mathbf{d}_{p_{1}}\left(0, \frac{x_{t}^{j}}{p_{0}} \right) \label{2t2d3}
\end{eqnarray}

\section{Examples} \label{examples}

\subsection{CA rules derived from binary operators}

We consider CA rules for which $l+r+1=2$ and take, for example $l=0$, $r=1$ below. Thus, the two inputs $x_{t}^{j-1} \in S$ and $x_{t}^{j}\in S$ at time $t$ and neighboring locations $j-1$ and $j$ yield an output $x_{t+1}^{j} \in S$ at time $t+1$. Eq. (\ref{themap}) takes in this case the simple form 
\begin{equation}
x_{t+1}^{j}=\mathbf{d}_{p}(x_{t}^{j-1}+px_{t}^{j}, R) \equiv x_{t}^{j} * x_{t}^{j-1} 
\end{equation}
where we have introduced the notation $*$ to emphasize that $\mathbf{d}_{p}(x_{t}^{j-1}+px_{t}^{j}, R)$ acts as a binary operator $S^{2} \to S$. Thus, the integer $R \in [0,p^{2}-1]$ lists all possible binary operators (magmas) on the set $S$ \cite{PHYSAFRAC}. If we tabulate the output $x_{t+1}^{j}$ as a function of the inputs $x_{t}^{j}$ and $x_{t}^{j-1}$ in the rows and in the columns respectively, we obtain the Cayley table of the binary operator as \cite{PHYSAFRAC}
\begin{center}
\begin{tabular}{c|cccccc}
$\ * \ $ & $ 0 \ $ & $\ 1 \ $ &  $\   \ldots  \ $ & $ \ p-1 \ $ \\
\hline
$\ 0 $ & $ \mathbf{d}_{p}(0, R) \ $ & $\ \mathbf{d}_{p}(1, R) \ $ & $\   \ldots  \ $ & $ \ \mathbf{d}_{p}(p-1, R) \ $ \\
$\ 1 $ & $ \mathbf{d}_{p}(p, R) \ $ & $\ \mathbf{d}_{p}(p+1, R) \ $ &  $\   \ldots  \ $ &  $ \ \mathbf{d}_{p}(2p-1, R) \ $ \\
$\ \ldots $ & $ \ldots \ $ & $\ \ldots \ $  & $\   \ldots  \ $ &  $ \ \ldots \ $ \\
$\ p-2 $ & $ \mathbf{d}_{p}(p(p-2), R) \ $ & $\ \mathbf{d}_{p}(p(p-2)+1, R) \ $ & $\   \ldots  \ $ &  $ \ \mathbf{d}_{p}(p(p-1)-1, R) \ $ \\
$\ p-1 $ & $ \mathbf{d}_{p}(p(p-1), R) \ $ & $\ \mathbf{d}_{p}(p(p-1)+1, R) \ $ &  $\   \ldots  \ $ &  $ \ \mathbf{d}_{p}(p^{2}-1, R) \ $ \\
\end{tabular}
\end{center}

Among these rules, most interesting ones are those with a group structure. Let us consider, for example, rules with Wolfram code
\begin{equation} 
R=\sum_{n=0}^{p^{2}-1}p^{n}\mathbf{d}_{p}\left(0,\mathbf{d}_{p}\left(0, n\right)+\mathbf{d}_{p}\left(1, n\right) \right) \label{Wcy}
\end{equation}
For these rules, from Eq. (\ref{themap}), we have \cite{PHYSAFRAC}
\begin{eqnarray}
x_{t+1}^{j}&=&\mathbf{d}_{p}\left(x_{t}^{j-1}+px_{t}^{j}, R\right)=\mathbf{d}_{p}\left(x_{t}^{j-1}+px_{t}^{j}, \sum_{n=0}^{p^{2}-1}p^{n}\mathbf{d}_{p}\left(0,\mathbf{d}_{p}\left(0, n\right)+\mathbf{d}_{p}\left(1, n\right) \right)\right) \nonumber \\
&=& \mathbf{d}_{p}\left(0, \mathbf{d}_{p}\left(0,\mathbf{d}_{p}\left(0, x_{t}^{j-1}+px_{t}^{j}\right)+\mathbf{d}_{p}\left(1, x_{t}^{j-1}+px_{t}^{j}\right) \right)\right) \nonumber \\
&=&
\mathbf{d}_{p}\left(0,\mathbf{d}_{p}\left(0, x_{t}^{j-1}\right)+\mathbf{d}_{p}\left(0, x_{t}^{j}\right) \right)=\mathbf{d}_{p}\left(0, x_{t}^{j-1}+x_{t}^{j} \right)
\end{eqnarray}
We now observe that the integers $x_{t}^{j-1}$ and $x_{t}^{j}$ in $S$ have a cyclic group structure $C_{p}$ under the action of the operator
\begin{equation}
\mathbf{d}_{p}\left(0, x_{t}^{j}+x_{t}^{j-1} \right)  \label{cyclope}
\end{equation}
Indeed, if we tabulate $x_{t+1}^{j}$ as a function of $x_{t}^{j}$ in the rows of the table and $x_{t}^{j-1}$ in the columns, we have \cite{CHAOSOLFRAC}
\begin{center}
\begin{tabular}{c|cccccc}
%\hline 
$\ \mathbf{d}_{p}\left(0, x_{t}^{j-1}+x_{t}^{j} \right)  \ $ & $ 0 \ $ & $\ 1 \ $ & $\ 2 \ $ & $\   \ldots  \ $ & $ \ p-2 \ $ & $ \ p-1 \ $ \\
\hline
$\ 0 $ & $ 0 \ $ & $\ 1 \ $ & $\ 2 \ $ & $\   \ldots  \ $ & $ \ p-2 \ $ & $ \ p-1 \ $ \\
$\ 1 $ & $ 1 \ $ & $\ 2 \ $ & $\ 3 \ $ & $\   \ldots  \ $ & $ \ p-1 \ $ & $ \ 0 \ $ \\
$\ \ldots $ & $ \ldots \ $ & $\ \ldots \ $ & $ \ \ldots \ $ & $\   \ldots  \ $ & $ \ \ldots \ $ & $ \ \ldots \ $ \\
$\ p-2 $ & $ p-2 \ $ & $\ p-1 \ $ & $\ 0 \ $ & $\   \ldots  \ $ & $ \ p-4 \ $ & $ \ p-3 \ $ \\
$\ p-1 $ & $ p-1 \ $ & $\ 0 \ $ & $\ 1 \ $ & $\   \ldots  \ $ & $ \ p-3 \ $ & $ \ p-2 \ $ \\
%\hline
\end{tabular}
\end{center}
which corresponds to the Cayley table of the finite cyclic group $C_{p}$ \cite{Gallian}. 
The invariance under addition modulo $p$ of CA with this structure \cite{VGM2} comes from the very fact that the digit function is a group homomorphism sending the integers $\mathbb{Z}$ to the set of residue classes $\mathbb{Z}/p\mathbb{Z}$ \cite{CHAOSOLFRAC}. In a previous work \cite{VGM3}, CA with this structure were suggested as templates for the design of the most complex (Class 4) CA.

If we now take $p_{0}=p_{1}=2$ we can, e.g, construct the direct sum $C_{2}\oplus C_{2}$ from Eqs. (\ref{p0p1gra}) and (\ref{Wcy}) by taking $A_{2}^{(1,0)}=A_{2}^{(1,1)}=\sum_{n=0}^{2^{2}-1}2^{n}\mathbf{d}_{2}\left(0,\mathbf{d}_{2}\left(0, n\right)+\mathbf{d}_{2}\left(1, n\right) \right)=6$, $r=0$ and $l=1$, i.e a CA rule of range $\rho=2$. Eq. (\ref{p0p1gra}) reduces in this case to
\begin{equation}
x_{t+1}^{j}=\mathbf{d}_{2}\left(0, x_{t}^{j}+x_{t}^{j-1} \right)+2\mathbf{d}_{2}\left(0,\mathbf{d}_{2}\left(0, \frac{x_{t}^{j}}{2} \right)+\mathbf{d}_{2}\left(0, \frac{x_{t}^{j-1}}{2} \right) \right) \label{p0p1gracy}
\end{equation}
And we now obtain the table
\begin{center}
\begin{tabular}{c|cccc}
$*$ & $ 0 \ $ & $\ 1 \ $ & $\   2 \ $ & $ \ 3 \ $\\
\hline
$\ 0 $ & $\ 0 \ $ & $\ 1 \ $ & $\ 2 \ $ & $\ 3 \ $\\
$\ 1 $ & $\ 1 \ $ & $\ 0 \ $ & $\  3 \ $ & $\ 2 \ $\\
$\ 2 $ & $\ 2 \ $ & $\ 3 \ $ & $ \ 0 \ $ & $\ 1 \ $\\
$\ 3 $ & $\ 3 \ $ & $\ 2 \ $ & $ \ 1 \ $ & $\ 0 \ $\\
\end{tabular}
\end{center}
which corresponds to the Cayley table of the direct sum $C_{2}\oplus C_{2}$ which is isomorphic to Klein's four group $V_{4}$. The CA so obtained has a diagram given by Fig. \ref{2times2} a). 

\begin{figure*} 
\includegraphics[width=1.0 \textwidth]{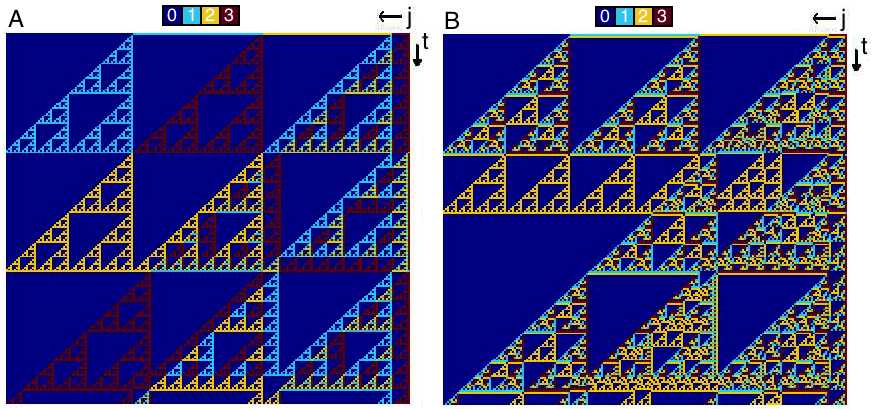}
\caption{\scriptsize{Spatiotemporal evolution of $x_{t}^{j}$ for the CA rules given by Eqs. (\ref{p0p1gracy}) (A) and (\ref{cyclo4}) (B) starting from an initial condition $x_{0}^{j}=\mathbf{d}_{4}(3,N_{s}-j)$ on a ring of 200 sites and for 200 time steps.}} \label{C2C2andC4}
\end{figure*}

In Fig. \ref{C2C2andC4} A the spatiotemporal evolution of $x_{t}^{j}$ obtained from Eq. (\ref{p0p1gracy}), is plotted for an initial condition $x_{0}^{j}=\mathbf{d}_{4}(3,N_{s}-j)$ on a ring of 200 sites and for 200 time steps. The latter initial condition is chosen to investigate the CA evolution starting from homogeneous patches that cover the whole range of possible values for $x_{0}^{j} \in [0,3]$. In this way, closed substructures involving subsets of the rule are revealed: The figure displays Sierpinski-like triangles of different colors that are clearly separated against the zero value of the background. These structures are the spatiotemporal realization of the subgroups of the direct sum $C_{2}\oplus C_{2}$, which correspond to the subsets $\{0,1\}$, $\{0,2\}$ and $\{0,3\}$ (Fig. \ref{C2C2andC4} A).

It is useful to compare the above CA rule with the one with $p=4$ with the structure of the cyclic group $C_{4}$. From Eq. (\ref{themap}) it is given by
\begin{equation}
x_{t+1}^{j}=\mathbf{d}_{4}\left(0, x_{t}^{j}+x_{t}^{j-1} \right)  \label{cyclo4}
\end{equation}
If we again tabulate the possible outputs $x_{t+1}^{j}$ of the CA by listing $x_{t}^{j+1}$ in the rows and $x_{t}^{j}$ in the columns, we now obtain 
\begin{center}
\begin{tabular}{c|cccc}
%\hline 
$\mathbf{d}_{4}\left(x_{t}^{j}+4x_{t}^{j+1}, R \right)  \ $ & $ 0 \ $ & $\ 1 \ $ & $\   2 \ $ & $ \ 3 \ $\\
\hline
$\ 0 $ & $\ 0 \ $ & $\ 1 \ $ & $\ 2 \ $ & $\ 3 \ $\\
$\ 1 $ & $\ 1 \ $ & $\ 2 \ $ & $\  3 \ $ & $\ 0 \ $\\
$\ 2 $ & $\ 2 \ $ & $\ 3 \ $ & $ \ 0 \ $ & $\ 1 \ $\\
$\ 3 $ & $\ 3 \ $ & $\ 0 \ $ & $ \ 1 \ $ & $\ 2 \ $\\
%\hline
\end{tabular}
\end{center}
After a little algebra it is seen that Eq. (\ref{cyclo4}) is decomposed, from Eqs. (\ref{atp1gen})  and (\ref{couply0}), as
\begin{eqnarray}
x_{t+1}^{j}
&=&\mathbf{d}_{2}\left(0,y_{t}^{(1,0)j}+y_{t}^{(1,0),j-1}\right)+\nonumber \\
&&+2\sum_{n=0}^{2}\mathbf{d}_{2}\left(y_{t}^{(1,1),j-1}+2y_{t}^{(1,1),j}, 6\right)\mathbf{d}_{2}\left(n-(y_{t}^{(1,0),j-1}+2y_{t}^{(1,0),j}),1\right)\nonumber \\
&&+2\mathbf{d}_{2}\left(y_{t}^{(1,1),j-1}+2y_{t}^{(1,1),j},9\right)
\mathbf{d}_{2}\left(3-(y_{t}^{(1,0),j-1}+2y_{t}^{(1,0),j}),1\right) \nonumber 
\end{eqnarray}
from which we observe that the rule is of the type of Fig. \ref{2times2} b), with $A_{n,p_{1}}^{(1,1)}=A_{n,2}^{(1,1)}=6$ for $n=0,1,2$ and $A_{n,2}^{(1,1)}=9$ for $n=3$. 

The spatiotemporal evolution of the CA rule given by Eq. (\ref{cyclo4}), in Fig. \ref{C2C2andC4} B, for the same initial condition, system size and number of time steps as in Fig. \ref{C2C2andC4} A. We now observe that only certain triangles involving the elements $\{0, 2\}$ (the only subgroup of $C_{4}$, which is isomorphic to $C_{2}$) can spatially separated from the others, where all symbols in $S$ are mixed. This is in contrast with (Fig. \ref{C2C2andC4} A) in which, as we have seen above, different triangular structures formed by the subsets $\{0,1\}$, $\{0,2\}$ and $\{0,3\}$ are separated. From this we see once more that \emph{the spatiotemporal evolution of the CA reveals closed substructures within the algebraic operator that specifies the rule.}

\subsection{Decomposable CA models based on the majority rule}

An important totalistic rule is the majority rule \cite{Vichniac,Tchuente,Goles,GolesBOOK,Ilachinski,Agur,Agur2}, which is given by 
\begin{equation}
\label{majo}
x_{t+1}^{j}=H\left(-\frac{1}{2}+\frac{1}{2\xi+1}\sum_{k=-\xi}^{\xi}x_{t}^{j+k} \right) 
\end{equation}
where $H(x)$ is the Heaviside function ($H(x)=0$ for $x <0$, $H(0)=\frac{1}{2}$ and $H(x)=1$ for $x >0$). This is a symmetric rule with $l=r=\xi$ with $p=2$ which returns, at each location $j$, the value '0' or '1' that occurs more frequently within the neighborhood of radius $\xi$ and $2\xi+1$ sites. It can be  seen that the majority rule is equivalently described by Eq. (\ref{themap2T}) as
\begin{equation}
x_{t+1}^{j}=\sum_{n=\left \lfloor \frac{2\xi+3}{2} \right \rfloor}^{2\xi+1}\mathbf{d}_{2}\left(n-\sum_{k=-\xi}^{\xi}x_{t}^{j+k} , 1 \right) \label{totama1}
\end{equation} 
or, alternatively, by Eq. (\ref{themap}) as
\begin{equation}
x_{t+1}^{j}=\mathbf{d}_{2}\left(\sum_{k=-\xi}^{\xi}x_{t}^{j+k} , 2^{2(\xi+1)}-2^{\left \lfloor (2\xi+3)/2 \right \rfloor} \right)
\label{totama2}
\end{equation}
Starting from an arbitrary initial condition consisting of `0's and `1's, this rule attains, after a short transient, a fixed point where domains are formed with no less than $\xi+1$ consecutive zeros or $\xi+1$ ones. For a system size of $N_{s}$ sites there are 
\begin{equation}
\mathcal{F}(N_{s}, \xi)=2+\sum_{\ell=1}^{\left \lfloor \frac{N_{s}}{2(\xi+1)} \right \rfloor}\frac{2N_{s}}{N_{s}-2\ell \xi}{N_{s}-2\ell \xi \choose 2\ell } \label{AgurAgur}
\end{equation}
such fixed points \cite{Agur}. Since there are $2^{N_{s}}$ possible initial conditions for a ring of $N_{s}$ sites, it is clear that, since, $\mathcal{F}(N_{s}, \xi) < 2^{N_{s}}$, the phase space irreversibly contracts so that the (dimensionless) entropy change involved in attaining any fixed point is 
\begin{equation}
\Delta S=\ln \left( \frac{\mathcal{F}(N_{s}, \xi)}{2^{N_{s}}}\right) < 0
\end{equation}
That the entropy is 'spontaneously' lowered in the CA evolution is not surprising if one thinks that the majority rule, as applied locally at each site $j$, acts as a Maxwell demon \cite{Sagawa}, by locally setting the value $x_{t+1}^{j}$  to the value that most often appears within the neighborhood of radius $\xi$ centered at $j$ at time $t$.

\begin{figure*} 
\includegraphics[width=1.0 \textwidth]{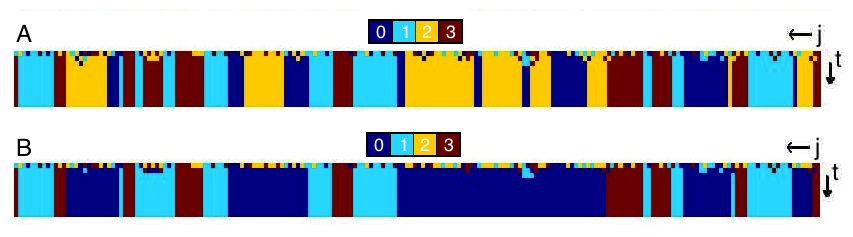}
\caption{\scriptsize{A. Spatiotemporal evolution of $x_{t}^{j}$ obtained from Eqs. (\ref{ama}) to (\ref{amaC}). B. Spatiotemporal evolution of $x_{t}^{j}$ obtained from Eqs. (\ref{amb}) to (\ref{ambC}) for $\xi_{0}=3$, $\xi_{1}=1$. In both cases the same  parameter values $\xi_{0}=3$, $\xi_{1}=1$ and same random initial condition are used, and 20 iteration steps are shown for a ring with $N_{s}=200$ sites.}} \label{majo1Dr}
\end{figure*}

We can now generalize the majority rule to domains with site values not restricted to be zero or one. For, if we consider a graded rule with diagram given by Fig. \ref{gsgp1} in which all layers are majority rules we have
\begin{equation}
x_{t+1}^{j} =\sum_{h=0}^{N-1}2^{h}y_{t+1}^{(1,h),j} \label{amaGE}
\end{equation}
where
\begin{equation}
y_{t+1}^{(1,h),j}=H\left(-\frac{1}{2}+\frac{1}{2\xi_{h}+1}\sum_{k=-\xi_{h}}^{\xi_{h}}y_{t}^{(1,h),j+k} \right)  \qquad h=0,1,\ldots, N-1 \label{amaBGE}
\end{equation}
and
\begin{equation}
y_{t}^{(1,h),j} \equiv \mathbf{d}_{2}\left(0, \frac{x_{t}^{j}}{2^{h}} \right) \qquad h=0,1,\ldots, N-1 \label{amaCGE}
\end{equation}
Since the layers are independent, the CA attains again a fixed point involving values in the set $\{0,1,2,\ldots,2^{N}-1\}$. There is now a total number of fixed points given by
\begin{equation}
\prod_{h=0}^{N-1}\mathcal{F}(N_{s}, \xi_{h})=\prod_{h=0}^{N-1}\left[2+\sum_{\ell=1}^{\left \lfloor \frac{N_{s}}{2(\xi_{h}+1)} \right \rfloor}\frac{2N_{s}}{N_{s}-2\ell \xi_{h}}{N_{s}-2\ell \xi_{h} \choose 2\ell } \right] \label{AgurAgurH}
\end{equation}
and, therefore, since the multiplicity of fixed points factorizes, the entropy change in attaining any of the fixed points starting from an arbitrary initial condition is the additive sum of the contributions from all independent layers
\begin{equation}
\Delta S=\ln \left( \frac{\prod_{h=0}^{N-1}\mathcal{F}(N_{s}, \xi_{h})}{2^{N_{s}}}\right)=\sum_{h=0}^{N-1}\ln \left( \frac{\mathcal{F}(N_{s}, \xi_{h})}{2^{N_{s}}}\right) < 0 \label{entrogra}
\end{equation}
In the case $N=1$ Eqs. (\ref{ama}) to (\ref{amaC}) reduce to Eq. (\ref{totama2}). Let us consider the case $N=2$ so that the model has diagram as in Fig. \ref{arbitrary} a). Eqs. (\ref{amaGE}) to (\ref{amaCGE}) reduce in this case to
\begin{eqnarray}
x_{t+1}^{j} &=&y_{t+1}^{(1,0),j}+2y_{t+1}^{(1,1),j} \label{ama} \\
y_{t+1}^{(1,0),j}&=&H\left(-\frac{1}{2}+\frac{1}{2\xi_{0}+1}\sum_{k=-\xi_{0}}^{\xi_{0}}y_{t}^{(1,0),j+k} \right)   \\
y_{t+1}^{(1,1),j}&=&H\left(-\frac{1}{2}+\frac{1}{2\xi_{1}+1}\sum_{k=-\xi_{1}}^{\xi_{1}}y_{t}^{(1,1),j+k} \right) \label{amaB} \\
y_{t}^{(1,0),j} &=& \mathbf{d}_{2}\left(0, x_{t}^{j} \right) \qquad \qquad y_{t}^{(1,1),j} = \mathbf{d}_{2}\left(0, \frac{x_{t}^{j}}{2} \right)
\label{amaC}
\end{eqnarray}
In Fig. \ref{majo1Dr}A the spatiotemporal evolution of $x_{t}^{j}$, obtained from Eqs. (\ref{ama}) to (\ref{amaC}) is shown for $\xi_{0}=3$, $\xi_{1}=1$ for a random initial condition $x_{0}^{j}$, and 20 iteration steps. We see that the system reaches a spatial fixed point where the different possible values of $S$, ($\{0,1,2,3\}$ in this case) arrange in spatial domains. Note that domains with value $x_{t}^{j}=2$ are contained within domains of value $x_{t}^{j}=0$ while those with value $x_{t}^{j}=3$ are contained within domains with value $x_{t}^{j}=1$. The reason is that $\mathbf{d}_{2}(0,2)=0$ and $\mathbf{d}_{2}(0,3)=1$ for a nonzero value $y_{t}^{(1,0)}=1$ of the lifted layer: If the lifted layer is nonzero, for $x_{t}^{j}=2$ (resp. $x_{t}^{j}=3$), the ground layer has value $y_{t}^{(1,0)}=0$ (resp. $y_{t}^{(1,0)}=1$). Note that although the layers are \emph{independent}, if the lifted layer has value zero at some $j$, the ground layer is, anyway, spatially correlated with the ground layer of the neighboring sites where the lifted layer is nonzero. This is reflected in the $x_{t}^{j}$ values. 

The graded CA rule can now be exploited to construct a CA model with coupled layers so as  to filter out some specific values of $S$ present in the initial condition. Let us consider again $N=2$ as in the model above, but now let us use the majority rule in the ground layer to influence the lifted layer, following the diagram of Fig. \ref{arbitrary} b). Specifically, we consider now the following model
\begin{eqnarray}
x_{t+1}^{j} &=&y_{t+1}^{(1,0),j}+2y_{t+1}^{(1,1),j} \label{amb} \\
y_{t+1}^{(1,0),j}&=&H\left(-\frac{1}{2}+\frac{1}{2\xi_{0}+1}\sum_{k=-\xi_{0}}^{\xi_{0}}y_{t}^{(1,0),j+k} \right)  \label{ambB} \\
y_{t+1}^{(1,1),j}&=&H\left(-\frac{1}{2}+\frac{1}{2\xi_{1}+1}\sum_{k=-\xi_{1}}^{\xi_{1}}y_{t}^{(1,1),j+k} \right)H\left(-\frac{1}{2}+\frac{1}{2\xi_{0}+1}\sum_{k=-\xi_{0}}^{\xi_{0}}y_{t}^{(1,0),j+k} \right)  \nonumber \\ && \label{amb2B} \\
y_{t}^{(1,0),j} &=& \mathbf{d}_{2}\left(0, x_{t}^{j} \right) \qquad \qquad y_{t}^{(1,1),j} = \mathbf{d}_{2}\left(0, \frac{x_{t}^{j}}{2} \right)
\label{ambC}
\end{eqnarray}
The lifted layer is now coupled to the ground layer and the model can be understood in terms of the simpler model of Eqs. (\ref{ama}) to (\ref{amaC}). Indeed, in those domains where $H\left(-\frac{1}{2}+\frac{1}{2\xi_{0}+1}\sum_{k=-\xi_{0}}^{\xi_{0}}y_{t}^{(1,0),j+k} \right)=1$ the model reduces to Eqs. (\ref{ama}) to (\ref{amaC}) and in those domains where 
$H\left(-\frac{1}{2}+\frac{1}{2\xi_{0}+1}\sum_{k=-\xi_{0}}^{\xi_{0}}y_{t}^{(1,0),j+k} \right)=0$ the model collapses to the simple majority rule, Eq. (\ref{majo}). As a consequence of this, it is readily observed that the value $x_{t}^{j}=2$ cannot occur in the trajectory of the CA and can only be present in the initial condition.

In Fig. \ref{majo1Dr}B the spatiotemporal evolution of $x_{t}^{j}$, obtained from Eqs. (\ref{amb}) to (\ref{ambC}) is shown for the same parameter values and initial condition as in In Fig. \ref{majo1Dr}A. We see that the system reaches a spatial fixed point that is similar to the one of Fig. \ref{majo1Dr}A with the exception that now $x_{t}^{j} \in \{0,1,3\}$ since the value $x_{t}^{j}=2$ is not possible in the trajectory. As a consequence of having filtered out the value $x_{t}^{j}=2$ through the coupling among layers, the entropy change from the initial condition till any of the fixed points is reached is \emph{lower} than in the graded rule case (i.e. Eq. (\ref{entrogra} with $N=2$): There is a smaller number of fixed points because the value $x_{t}^{j}=2$ is not possible in these, as a consequence of the coupling of the lifted layer to the ground layer. Thus, if we evaluate the entropy change with Eq. (\ref{entrogra}) without taking into account the coupling, we would overestimate the entropy change. \emph{That entropy is lowered through correlations shows that entropy is no longer equal to the mere addition of the independent layers}. Therefore, this constitutes a dynamical system in which the methods of nonextensive statistics \cite{Tsallis,VGMStat} and superstatistics \cite{Beck} may be used. In \cite{PNAS2,VGMmacroion} physical examples (both in and out of equilibrium) where interactions lead to a non-additive entropy (connected to a reduction in the available phase space) are given.

\begin{figure*} 
\includegraphics[width=1.0 \textwidth]{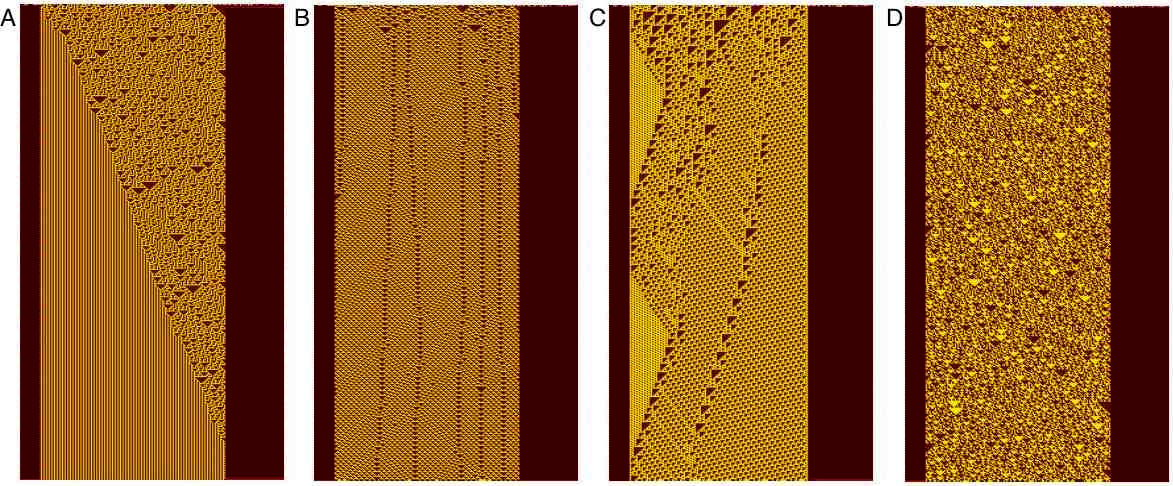}
\caption{\scriptsize{Spatiotemporal evolution of $x_{t}^{j}$ obtained from Eqs. (\ref{ambb}) to (\ref{ambbC})  and for $R=30$ (A), $R=54$ (B), $R=110$ (C) and $R=150$ (D). In all cases, $N_{s}=200$, $\xi= 30$ and $400$ time steps are shown starting from a random initial condition that is the same in all cases. Time flows from top to bottom and $j$ increases from right to left in each panel.}} \label{majotherules}
\end{figure*}

The dynamical behavior of any of the domains can be specified separately. For example, in the following model
\begin{eqnarray}
x_{t+1}^{j} &=&y_{t+1}^{(1,0),j}+2y_{t+1}^{(1,1),j} \label{ambb} \\
y_{t+1}^{(1,0),j}&=&H\left(-\frac{1}{2}+\frac{1}{2\xi+1}\sum_{k=-\xi}^{\xi}y_{t}^{(1,0),j+k} \right)  \label{ambbB} \\
y_{t+1}^{(1,1),j}&=&\mathbf{d}_{2}\left(\sum_{k=-1}^{1}2^{k+r}y_{t}^{(1,1),j+k} , R  \right)H\left(-\frac{1}{2}+\frac{1}{2\xi+1}\sum_{k=-\xi}^{\xi}y_{t}^{(1,0),j+k} \right)\nonumber \\&&  \label{ambb2B} \\
y_{t}^{(1,0),j} &=& \mathbf{d}_{2}\left(0, x_{t}^{j} \right) \qquad \qquad y_{t}^{(1,1),j} = \mathbf{d}_{2}\left(0, \frac{x_{t}^{j}}{2} \right)
\label{ambbC}
\end{eqnarray}
the domains for which $y_{t}^{(1,0),j}=0$ (once a fixed point is attained) remain at the quiescent state $x_{t}^{j}=0$ forever. However, in those domains for which $y_{t}^{(1,0),j}=1$ ($t \to \infty$), $x_{t}^{j}=1+2y_{t+1}^{(1,1),j}$ from Eq. (\ref{ambb}) and $y_{t+1}^{(1,1),j}=\mathbf{d}_{2}\left(\sum_{k=-1}^{1}2^{k+r}y_{t}^{(1,1),j+k} , R  \right)$ from Eq. (\ref{ambb2B}). The latter corresponds to a Wolfram elementary CA with code $0\le R \le 255$ and, therefore, such Wolfram Boolean CA runs within the domains where the ground layer has value '1' taking values on the set $x_{t}^{j} \in \{1,3\}$.  In Fig.\ref{majotherules} the spatiotemporal evolution of $x_{t}^{j}$ obtained from Eqs. (\ref{ambb}) to (\ref{ambbC})  and for $R=30$ (A), $R=54$ (B), $R=110$ (C) and $R=150$ (D), is shown. In all cases, $N_{s}=200$, $\xi= 30$ and $400$ time steps are shown starting from a random initial condition that is the same in all cases. Time flows from top to bottom and $j$ increases from right to left in each panel. We observe that, because of interaction with the borders of the domain, the chaos present in Wolfram's CA rule 30 vanishes after ca. 380 time steps (panel A). Rules 54 and 110 (panels B and C) display long transients with unpredictable behavior, and the additive rule 150 yields a chaotic pattern confined to the domain. In all cases, there is a single domain where the interesting dynamics takes place coexisting with a domain in the quiescent state. All cases correspond to the class of models given by Fig. \ref{arbitrary} b).

\begin{figure} 
\includegraphics[width=0.7 \textwidth]{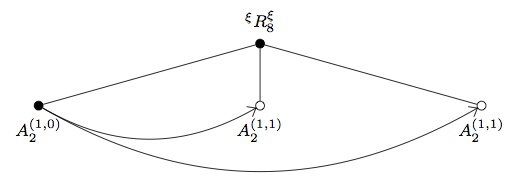}
\caption{\scriptsize{Diagram corresponding to the CA model for chimera states in \cite{EPL}, Eqs. (\ref{ambbchi}) to (\ref{ambbchiC}).
}} \label{chidia}
\end{figure}

The following is a CA model for chimera states that has been introduced in a recent article \cite{EPL} 
\begin{eqnarray}
x_{t+1}^{j} &=&y_{t+1}^{(1,0),j}+2y_{t+1}^{(1,1),j}+4y_{t+1}^{(1,2),j} \label{ambbchi} \\
y_{t+1}^{(1,0),j}&=&H\left(-\frac{1}{2}+\frac{1}{2\xi+1}\sum_{k=-\xi}^{\xi}y_{t}^{(1,0),j+k} \right)  \label{ambbchiB} \\
y_{t+1}^{(1,1),j}&=&\left[1-H\left(-\frac{1}{2}+\frac{1}{2\xi+1}\sum_{k=-\xi}^{\xi}y_{t}^{(1,1),j+k} \right) \right]\times \nonumber \\
&&\times \left[1-H\left(-\frac{1}{2}+\frac{1}{2\xi+1}\sum_{k=-\xi}^{\xi}y_{t}^{(1,0),j+k} \right)\right]  \label{ambbchiB} \\
y_{t+1}^{(1,2),j}&=&\mathbf{d}_{2}\left(\sum_{k=-1}^{1}2^{k+r}y_{t}^{(1,1),j+k} , 105  \right)H\left(-\frac{1}{2}+\frac{1}{2\xi+1}\sum_{k=-\xi}^{\xi}y_{t}^{(1,0),j+k} \right) \nonumber \\&&  \label{ambbchi2B} \\
y_{t}^{(1,0),j} &=& \mathbf{d}_{2}\left(0, x_{t}^{j} \right) \qquad y_{t}^{(1,1),j} = \mathbf{d}_{2}\left(0, \frac{x_{t}^{j}}{2} \right) \qquad y_{t}^{(1,2),j} = \mathbf{d}_{2}\left(0, \frac{x_{t}^{j}}{4} \right) \nonumber \\
&& \label{ambbchiC}
\end{eqnarray}
It has diagram given by Fig. \ref{chidia}, where the ground layer, which is the majority rule, is a layer CA that influences the remaining lifted layers. The model works on eight symbols $p=8=2^{3}$ and decomposes into layers working on two symbols each (Boolean layers).

\begin{figure*} 
\includegraphics[width=1.0 \textwidth]{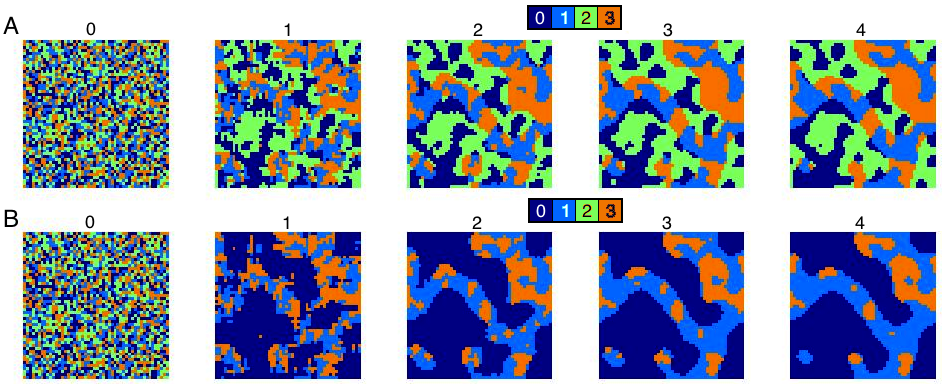}
\caption{\scriptsize{A. Spatiotemporal evolution of $x_{t}^{j}$ obtained from Eqs. (\ref{ama2D}) to (\ref{amaC2D}). B. Spatiotemporal evolution of $x_{t}^{j}$ obtained from Eqs. (\ref{ama2Db}) to (\ref{amaC2Db}). In both cases the same parameter values $\xi_{0}=3$, $\xi_{1}=1$ and same random initial condition are used, and 4 iteration steps (indicated over the panels) are shown for a surface with $N_{s}\times N_{s}=50\times 50$ sites.}} \label{majo2D}
\end{figure*}

As we have mentioned above, all above models can be generalized to any dimension and/or neighborhood by introducing more spatial coordinates. Let $i$ be an index that besides $j$ runs on a square surface under periodic boundary conditions. For example, for two dimensions and square Moore neighborhoods with $2\xi_{h}+1$ side (for the layer $h=0,1\ldots, N-1$) we have that Eqs. (\ref{amaGE}) to (\ref{amaCGE}) generalize as
\begin{equation}
x_{t+1}^{i,j} =\sum_{h=0}^{N-1}2^{h}y_{t+1}^{(1,h),i,j} \label{ama2D}
\end{equation}
where
\begin{equation}
y_{t+1}^{(1,h),i,j}=H\left(-\frac{1}{2}+\frac{1}{(2\xi_{h}+1)^{2}}\sum_{m=-\xi_{h}}^{\xi_{h}}\sum_{k=-\xi_{h}}^{\xi_{h}}y_{t}^{(1,h),i+k,j+m} \right)  \qquad h=0,1,\ldots, N-1 \label{amaBgene2D}
\end{equation}
and
\begin{equation}
y_{t}^{(1,h),i,j} \equiv \mathbf{d}_{2}\left(0, \frac{x_{t}^{i,j}}{2^{h}} \right) \qquad s=0,1,\ldots, N-1 \label{amaC2D}
\end{equation}
If we take $N=2$ in these expressions, we obtain the corresponding graded rule in two dimensions. We obtain
\begin{eqnarray}
x_{t+1}^{i,j} &=&y_{t+1}^{(1,0),i,j}+2y_{t+1}^{(1,1),i,j} \label{ama2D} \\
y_{t+1}^{(1,0),i,j}&=&H\left(-\frac{1}{2}+\frac{1}{(2\xi_{0}+1)^{2}}\sum_{m=-\xi_{0}}^{\xi_{0}}\sum_{k=-\xi_{0}}^{\xi_{0}}y_{t}^{(1,0),i+k,j+m} \right)   \\
y_{t+1}^{(1,1),i,j}&=&H\left(-\frac{1}{2}+\frac{1}{(2\xi_{1}+1)^{2}}\sum_{m=-\xi_{1}}^{\xi_{1}}\sum_{k=-\xi_{1}}^{\xi_{1}}y_{t}^{(1,1),i+k,j+m} \right) \label{amaB2D} \\
y_{t}^{(1,0),i,j} &=& \mathbf{d}_{2}\left(0, x_{t}^{i,j} \right) \qquad \qquad y_{t}^{(1,1),i,j} = \mathbf{d}_{2}\left(0, \frac{x_{t}^{i,j}}{2} \right)
\label{amaC2D}
\end{eqnarray}
These equations generalize Eqs. (\ref{ama}) to (\ref{amaC}) to two dimensions. In Fig. \ref{majo2D}A the spatiotemporal evolution of $x_{t}^{i,j}$ obtained from Eqs. (\ref{ama2D}) to (\ref{amaC2D}) is shown for a surface with $N_{s}\times N_{s}=50\times 50$ sites, $\xi_{0}=3$, $\xi_{1}=1$ and a random initial condition. Four iteration steps are shown. After a short transient a 2D spatial fixed point is reached which is analogous in 2D to the one found for 1D in Fig. \ref{majo1Dr}A.

Eqs. (\ref{amb}) to (\ref{ambC}) generalize to 2D as
\begin{eqnarray}
x_{t+1}^{i,j} &=&y_{t+1}^{(1,0),i,j}+2y_{t+1}^{(1,1),i,j} \label{ama2Db} \\
y_{t+1}^{(1,0),i,j}&=&H\left(-\frac{1}{2}+\frac{1}{(2\xi_{0}+1)^{2}}\sum_{m=-\xi_{0}}^{\xi_{0}}\sum_{k=-\xi_{0}}^{\xi_{0}}y_{t}^{(1,0),i+k,j+m} \right)   \\
y_{t+1}^{(1,1),i,j}&=&H\left(-\frac{1}{2}+\frac{1}{(2\xi_{1}+1)^{2}}\sum_{m=-\xi_{1}}^{\xi_{1}}\sum_{k=-\xi_{1}}^{\xi_{1}}y_{t}^{(1,1),i+k,j+m} \right)\times \nonumber \\
&& \times H\left(-\frac{1}{2}+\frac{1}{(2\xi_{0}+1)^{2}}\sum_{m=-\xi_{0}}^{\xi_{0}}\sum_{k=-\xi_{0}}^{\xi_{0}}y_{t}^{(1,0),i+k,j+m} \right) \label{amaB2Db} \\
y_{t}^{(1,0),i,j} &=& \mathbf{d}_{2}\left(0, x_{t}^{i,j} \right) \qquad \qquad y_{t}^{(1,1),i,j} = \mathbf{d}_{2}\left(0, \frac{x_{t}^{i,j}}{2} \right)
\label{amaC2Db}
\end{eqnarray}
In Fig. \ref{majo2D}B we show the spatiotemporal evolution of $x_{t}^{i,j}$ obtained from Eqs. (\ref{ama2Db}) to (\ref{amaC2Db}) for a surface with $N_{s}\times N_{s}=50\times 50$ sites, $\xi_{0}=3$, $\xi_{1}=1$ and a random initial condition that is the same as in Fig. \ref{majo2D}A. After a short transient a 2D spatial fixed point is reached which is analogous in 2D to the one found for 1D in Fig. \ref{majo1Dr}B. Note the absence of the value $x_{t}^{i,j}=2$ in the trajectory of the CA. Interestingly, this slightly affects the distribution of the value $x_{t}^{i,j}=3$ when compared to Fig. \ref{majo2D}A. This is due to the fact that values $x_{t}^{j}=2$ and $x_{t}^{j}=3$ in the latter figure are correlated in the lifted layer with a value $y_{t+1}^{(1,1),i,j}=1$. Thus, removing the value $x_{t}^{j}=2$ as in Fig. \ref{majo2D}B causes that now a value $y_{t+1}^{(1,1),i,j}=1$ in the lifted layer is surrounded by more zeros, as it is in Fig. \ref{majo2D}A. This means that if regions with value $x_{t}^{j}=2$ (green) are absent as in Fig. \ref{majo2D}B, regions with value $x_{t}^{j}=3$ (orange) are smaller, compared to those of Fig. \ref{majo2D}A. This rather subtle effect is thus explained through the correlations that are clearly revealed through the construction of the CA model. We note that clustered structures implying formation of domains  arise naturally in many physical experimental systems under global constraints (see e.g. \cite{Conti}) and (in an oscillatory fashion) in nonlinear electrochemical systems (see e.g. Fig. 2b in \cite{ChimeraSCH14a}). Subclustering ('clusters-within-clusters') has been found ubiquitous in parameter space in a modified version of the complex Ginzburg-Landau equation under nonlinear global coupling \cite{ChimeraMiethe,ChimeraOrlov}. Note that with the method presented here, we can generalize this sub-clustering to an arbitrary number of layers, with the possibility of modeling the individual and differentiated dynamical behavior of each substructure as well as its interactions and relationships to the rest.

\section{Conclusions}

In this article we have presented a diagrammatic approach for the study and design of sophisticated CA models, as well as a general framework to mathematically formulate and analyze these models. In a previous paper \cite{semipredo}, we presented the main idea of the $p$-decomposability of CA rules (whose consequences have been further explored here) and the way of constructing graded CA rules or of \emph{decomposing} complicated non-graded CA rules with coupled layers. The theory presented in that paper is here completed by establishing the means and the systematic approach to \emph{compose} non-graded CA rules, so that the couplings, correlations and interactions of complex systems can be systematically modeled. Here, the traditional tables containing all possible outputs for configurations of the CAs are no longer necessary, and a few integer parameters suffice to produce complex CA models with desired properties. 

We have presented several examples showing how CA rules can be endowed with useful algebraic structure and how the spatiotemporal evolution of the CA indeed reveals closed algebraic substructures hidden within the specification of the rule. We have also shown how the majority rule can be used as a building block to spatially separate structures and substructures in the form of homogeneous domains and subdomains and we have thus generalized the majority rule to an arbitrary number of symbols and layers. With the mathematical method presented in this paper we can, with help of the diagrams, construct CA models, with as many substructures as layers, systematically including domains within domains in a 'Matryoshka dolls' fashion by means of the generalized majority rule discussed in this manuscript. Since the bulk dynamics of the domains can be systematically designed, as suggested by Fig. \ref{majotherules} (save for the complicated interaction with the borders and/or surprising correlations that may appear through the topology), the approach presented constitutes a general mechanism for the emergence of complex forms ('organisms') constituted by distinct spatially extended parts with different dynamical behaviors and potentially different functions.

%\section*{References}

\end{document}